\documentclass[10pt,a4paper,twocolumn]{article}

\usepackage{graphicx}% Include figure files
\usepackage{dcolumn}% Align table columns on decimal point
\usepackage{bm}% bold math
\usepackage{textcomp}
\usepackage{color}
\usepackage[normalem]{ulem}
\usepackage{gensymb}
\usepackage{amsmath}
\usepackage{acronym}
\usepackage{authblk}

\definecolor{JLAQ}{rgb}{1,0,1}

\begin{document}
\bibliographystyle{ieeetr}
\title{Broadband Circular-to-Linear Polarization Conversion of Terahertz Waves Using Self-Complementary Metasurfaces}
\author[1]{Andrey Sayanskiy\thanks{a.sayanskiy@metalab.ifmo.ru}}
\author[2,3]{Sergei A. Kuznetsov\thanks{serge$\_$smith@ngs.ru}}
\author[2]{Daria S. Tanygina}
\author[4]{Juan P. del Risco}
\author[1]{Stanislav Glybovski\thanks{s.glybovski@metalab.ifmo.ru}}
\author[5]{Juan D. Baena\thanks{jdbaenad@unal.edu.co}}

\affil[1]{Department of Nanophotonics  and  Metamaterials,  ITMO  University,  St.  Petersburg 197101, Russia}
\affil[2]{Rzhanov Institute of Semiconductor Physics SB RAS, Novosibirsk Branch "TDIAM," Lavrentiev Avenue 2/1, Novosibirsk 630090, Russian Federation}
\affil[3]{Physics Department, Novosibirsk State University, Pirogova Street 2, 630090 Novosibirsk, Russian Federation}
\affil[4]{Universidad Sergio Arboleda, School of Exact Sciences and Engineering, 111711, Bogota, Colombia}
\affil[5]{Physics Department, Universidad Nacional de Colombia, Bogota 111321, Colombia}

\maketitle

\section{Abstract}
   In this work, we theoretically and experimentally study the conversion from a circularly polarized plane electromagnetic wave into a linearly polarized transmitted one using anisotropic self-complementary metasurfaces. For this purpose, a metasurface design operable at sub-terahertz frequencies is proposed and investigated in the range of 230$-$540 GHz. The metasurface is composed of alternating complementary rectangular patches and apertures patterned in an aluminum layer deposited on a thin polypropylene film. The term \textit{self-complementary} implies that the pattern is invariant with respect to its inversion (with some translation). Our study shows that the translational symmetry of the metasurface results in unusual and useful electromagnetic properties under illumination with circularly polarized radiation beams. In particular, alongside with broadband circular-to-linear conversion, the transmitted wave exhibits a frequency independent magnitude, while its polarization angle gradually changes with frequency that paves the way for new beam-splitting applications.

%%%%%%%%%%%%%%%%%%%%%%%%%%%%%%%%%%%%%%%%%%%%%%%%%%%%%%%%%%%%%%%%%%%%%
%% Start the main part of the manuscript here.
%%%%%%%%%%%%%%%%%%%%%%%%%%%%%%%%%%%%%%%%%%%%%%%%%%%%%%%%%%%%%%%%%%%%%
\section{Introduction}
Metasurfaces (MSs) are defined as dense two-dimensional arrays of subwavelength scatterers \cite{Holloway12,Capasso14,Minovich2015195,Review,Chen2016}. Unlike diffraction gratings, MSs behave as optically thin and continuous sheets supporting the required distributions of averaged surface currents \cite{Review}. The latter is achieved by designing the structure of constituent elements, i.e. MS unit cells containing metal or dielectric scatterers. 

Last decade, following a rapid progress in terahertz (THz) wave technologies stimulated by the development of high-efficiency radiation sources and detectors \cite{SongBook}, a noticeable interest is taken in THz MSs as a new paradigm in designing high-performance and compact quasi-optical devices for radiation manipulation. It is worth noting that the THz band, whose frequencies extend from 0.1 to 10 THz or in terms of wavelength from 30 {\textmu}m to 3 mm, appears to be very convenient from the technological standpoint as it enables involving inexpensive photolithographic and other micromachining techniques of micrometer accuracy for producing large-area MSs of high quality. Nowadays, the THz MSs are used as frequency-selective surfaces (FSSs) \cite{Chen2006,Tao09,Zhang13,Morits14,Wang16}, perfect absorbers \cite{Tao08,Grant11,Liu17,KuznetsovSciRep}, polarization converters \cite{Strikwerda,Euler,Liu15,Mo16}, focusing structures \cite{Kuznetsov2015,Wang2016,Xueqian,Chang17}, sensors \cite{Beruete2017}, and some others.
FSSs are usually implemented as periodically patterned single-layer metal sheets of a negligible thickness. Despite unavoidable reflectance \cite{Barlevy}, the advantages of FSSs are in their thinness and relatively simple lithographic fabrication.
FSSs have been of a high interest for a long time due to applications in microwave resonant beam-splitters for reflector antennas \cite{Arnaud} and transparent antenna radomes \cite{Radome}.  More recently, FSSs have been proposed for using as linear-to-circular polarization converters in the microwave \cite{Meander1}, terahertz \cite{Strikwerda,Euler} and optical \cite{Tharp06,Li2015} ranges. 
The unit cells of such metasurfaces should be polarized by orthogonal components of the incident electric field equally strong but with a $90\degree$ phase shift. This is usually achieved by mutual detuning of two orthogonally rotated resonant scatterers comprising each unit cell. Alternatively, excitation of two orthogonally polarized eigenmodes of the same scatterer resonating at slightly different frequencies can be employed \cite{Review,ComparisonFSS}. In both approaches the geometry of the FSS unit cells should be carefully adjusted to provide equality of magnitudes and $\pm 90\degree$ difference in phases of transmitted waves with orthogonal linear polarizations at the same frequency. For this purpose, in most previously reported FSSs a variation of more than one geometric parameter must be performed. 

Recently, \textit{self-complementary} metasurfaces have been shown to have unique and useful electromagnetic properties\cite{Nakata,Baena_zigzag,Baena_self}. 
Such MSs are a class of symmetric FSSs whose metal pattern is identical to their Babinet-inverted complement. In other words, with the metal elements replaced by apertures and vice versa, the whole pattern is transformed to itself. 

One of the earliest examples of self-complementary FSSs with a fourfold rotational symmetry is an isotropic checkerboard FSS\cite{McPhedran,Hangyo}. However, square conductors in this structure must necessarily approach a zero distance between their perfectly sharp corners, which produce a singularity. Otherwise, there are either electric contacts or gaps between the conductors, so that the structure is no more self-complementary and qualitatively changes its properties \cite{McPhedran,Hangyo}. For this reason, an isotropic and loss-less self-complementary MS is impractical. This issue was resolved by using resistive (lossy) isotropic checkerboard FSSs providing the frequency independent transmittance of 0.25 (either for linear or circular polarization) \cite{Nakata}. Later, this property of resistive isotropic self-complementary MSs was experimentally confirmed in the terahertz range \cite{Kitano2015}. Moreover, a THz switchable quarter-wave plate based on a modified checkerboard structure was reported\cite{Nakata18}.

The other type of self-complementary MSs with a translational symmetry has been studied e.g. in the works \cite{Beruete,Beruete2,Ortiz}. As opposed to checkerboard MSs, such structures do not suffer from the problem of corners and are much simpler in technological realization as do not necessitate using suplementary resistive elements in the MS design. 
It was theoretically and experimentally shown that any self-complementary MS of this type can be used as a linear-to-circular polarization converter at least at one frequency \cite{Baena_self}. The narrowband \cite{Baena_self} and a wideband \cite{Baena_zigzag} unit cells have been proposed and experimentally studied in the microwave range. Unlike other single-layer FSSs optimized to operate as polarization converters\cite{Tharp06,Euler,Li2015}, self-complementary structures provide a frequency-constant phase difference of $\pm 90\degree$ between transmission coefficients of orthogonally polarized plane waves. This property, granted by the symmetry of their patterns, drastically simplifies the design of optically thin polarization converters. 

It is worth highlighting that under circularly polarized illumination any self-complementary MS with a translational symmetry demonstrates frequency-independent and equal to 1/2 coefficients of co-polarization transmission (i.e. referred to the field transfer from the right-handed (RH) or left-handed (LH) circular polarization (CP) to the same state: RHCP-to-RHCP or LHCP-to-LHCP)\cite{Nakata}. However, as it will be shown in this work, the cross-polarization transmission for the circular polarization, though having a constant magnitude of 1/2, exhibits a frequency-dependent phase. We theoretically and experimentally demonstrate that this property can be used for circular-to-linear polarization conversion (CP-to-LP) with the output polarization angle gradually changing with frequency. For this purpose, we consider a resonant structure composed of alternating patches and complementary apertures and further investigate it in the frequency range of 230$-$540 GHz. 

\section{Theory}\label{Th}

In this section we analytically describe the electromagnetic response of an arbitrary anisotropic self-complementary FSS under illumination with a normally incident and circularly polarized plane wave.

\begin{figure}[h]
  \centering
  \includegraphics[width=0.5\textwidth]{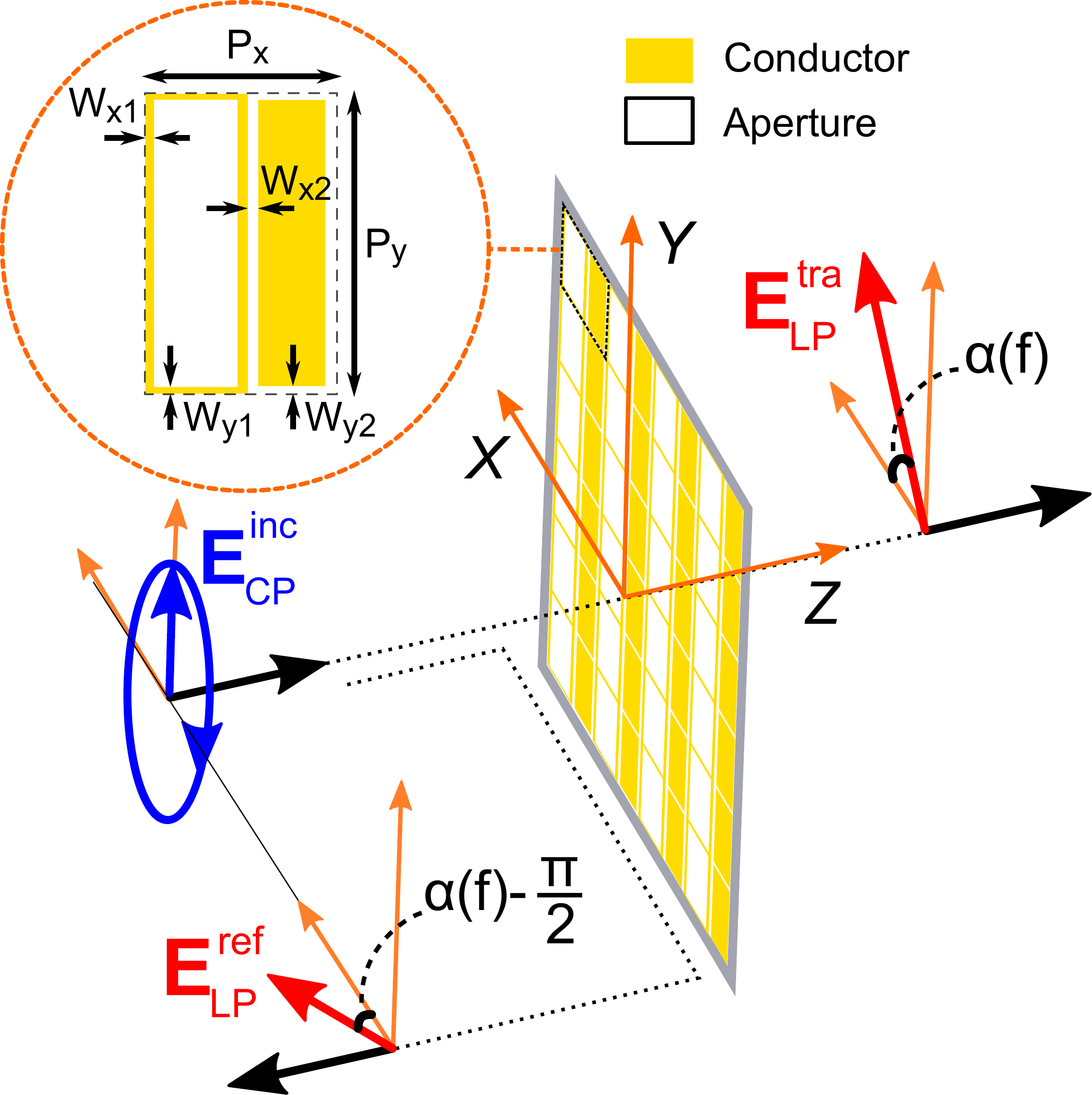} 
  \caption{Self-complementary metasurface with translational symmetry is illuminated by a right-handed circularly polarized plane wave propagating in the normal direction. The linearly-polarized waves are produced both in transmission and reflection. Inset shows a single unit cell comprised by a rectangular patch and a complementary aperture.}
  \label{Geometry}
\end{figure}

Let our metasurface have mirror symmetry with respect to $ZX$ and $YZ$ planes, where $z$ is the normal axis. An example of such a FSS pattern together with the chosen coordinate system is shown in Fig.~\ref{Geometry}.
Due to the selected symmetry, the transmission and reflection matrices for the linearly polarized field components in the current basis of $x$- and $y$-polarizations become diagonal\cite{DMITRIEV2011}:
\begin{equation}\label{MatrixTR}
\overline{\overline{T}}_{\text{LP}}=
 \begin{pmatrix}
      {t_x} & {0} \\
      {0} & {t_y}     
 \end{pmatrix};~~~
 \overline{\overline{R}}_{\text{LP}}=
 \begin{pmatrix}
      {r_x} & {0} \\
      {0} & {r_y}       
 \end{pmatrix}, 
\end{equation}
where $t_{x}=E_x^{\text{tra}}/E_x^{\text{inc}}$, $t_{y}=E_y^{\text{tra}}/E_y^{\text{inc}}$ and $r_{x}=E_x^{\text{ref}}/E_x^{\text{inc}}$, $r_{y}=E_y^{\text{ref}}/E_y^{\text{inc}}$ are the transmission and reflection coefficients for $x$- and $y$-polarization correspondingly. By the assumption  of anisotropy, $t_y \ne t_x$. Having determined these coefficients, the incidence of a wave with arbitrary polarization can be further analyzed. The same wave can be represented in the basis of orthogonal linear polarizations ($\textbf{e}_x$,$\textbf{e}_y$) and in the basis ($\textbf{e}_{\text{R}}$,$\textbf{e}_{\text{L}}$) of right- and left-handed circular ones: $\textbf{E}=E_x\textbf{e}_x+E_y\textbf{e}_y=E_{\text{R}}\textbf{e}_{\text{R}}+E_{\text{L}}\textbf{e}_{\text{L}}$, where $\textbf{e}_{\text{R}}=\left( 
\textbf{e}_x+i\textbf{e}_y \right)/\sqrt{2}$ and $\textbf{e}_{\text{L}}=\left( 
\textbf{e}_x-i\textbf{e}_y \right)/\sqrt{2}$, while $E_{\text{R}}$ and $E_{\text{L}}$ are the complex magnitudes of a right- and left-handed circular polarization components. The field components in ($\textbf{e}_{\text{R}}$,$\textbf{e}_{\text{L}}$) and ($\textbf{e}_x$,$\textbf{e}_y$) bases are related through the transformation matrix $\overline{\overline{M}}$:
\begin{equation}\label{TransformMatrix}
\begin{pmatrix}
      {E_R} \\
      {E_L}      
 \end{pmatrix}  =
\overline{\overline{M}}
 \begin{pmatrix}
      {E_x} \\
      {E_y}
 \end{pmatrix};~~~ 
\overline{\overline{M}}=
 \begin{pmatrix}
      {\frac{1}{\sqrt{2}}} & {\frac{-\text{i}}{\sqrt{2}}} \\
      {\frac{1}{\sqrt{2}}} & {\frac{\text{i}}{\sqrt{2}}} 
 \end{pmatrix}.
\end{equation}
Without loss of generality, further we consider in detail only the transmitted wave. The reflected wave can be analyzed in the same manner.
In the ($\textbf{e}_{\text{R}}$,$\textbf{e}_{\text{L}}$) basis the transmitted wave is calculated as 
\begin{equation}\label{TransmissionCP1}
 \begin{pmatrix}
      E_{\text{R}}^{\text{tra}} \\
      E_{\text{L}}^{\text{tra}}
 \end{pmatrix} = 
 \overline{\overline{T}}_{\text{CP}}
  \begin{pmatrix}
      E_{\text{R}}^{\text{inc}} \\
      E_{\text{L}}^{\text{inc}}
 \end{pmatrix};~
\overline{\overline{T}}_{\text{CP}} = \begin{pmatrix}
      t_{\text{RR}} & t_{\text{RL}} \\
      t_{\text{LR}} & t_{\text{LL}}
 \end{pmatrix}, 
\end{equation}
where $t_{\text{RR}}$ and $t_{\text{LL}}$ are the co-polar transmission coefficients for corresponding circular polarizations, while $t_{\text{RL}}$ and $t_{\text{LR}}$ are the cross-polar ones. In (\ref{TransmissionCP1}) $\overline{\overline{T}}_{\text{CP}}$ represents the transmission matrix referred to the ($\textbf{e}_{\text{R}}$,$\textbf{e}_{\text{L}}$) basis, which can be found from (\ref{MatrixTR}) and (\ref{TransformMatrix}) as
\begin{equation}\label{TransmissionCP2}
 \begin{gathered}
\overline{\overline{T}}_{\text{CP}}=\overline{\overline{M}}\cdot \overline{\overline{T}}_{\text{LP}} \cdot\overline{\overline{M}}^{-1}=
  \begin{pmatrix}
      {\frac{t_x+t_y}{2}} & {\frac{t_x-t_y}{2}} \\
      {\frac{t_x-t_y}{2}} & {\frac{t_x+t_y}{2}} \\ 
 \end{pmatrix}.
  \end{gathered}
\end{equation}
Previously, it was demonstrated that transmission coefficients of self-complementary MSs for orthogonal linear polarizations satisfy the condition\cite{Nakata,Baena_zigzag,Baena_self}: $t_x$+$t_y$=1. Moreover, $t_x$ and $t_y$ are phase-shifted by 90$^{\circ}$ at any frequency\cite{Baena_zigzag,Baena_self} and both the transmission phases $\arg(t_x)$ and $\arg(t_y)$ take values in the range from $-90^{\circ}$ to $90^{\circ}$. Therefore, the transmission coefficients for linear polarizations can be written in the form: $t_x=\cos(\phi_y \pm \pi/2) \exp{\text{i} (\phi_y\pm\pi/2)}$ and $t_y=\cos(\phi_y)\exp{\text{i}\phi_y}$, where $\phi_y$ is the transmission coefficient phase for $y$-polarization. Substituting the last expressions in (\ref{TransmissionCP2}) yields
\begin{equation}\label{TransmissionCPFinal}
\overline{\overline{T}}_{\text{CP}}=
  \begin{pmatrix}
      \frac{1}{2} & \frac{\exp{\text{i}(\pi+2\phi_y)}}{2}\\
      \frac{\exp{\text{i}(\pi+2\phi_y)}}{2} & \frac{1}{2}  
 \end{pmatrix}. 
\end{equation}
From (\ref{TransmissionCPFinal}) it is clearly seen that all magnitudes of co-polar ($t_{\text{RR}}$, $t_{\text{LL}}$) and cross-polar ($t_{\text{RL}}$, $t_{\text{LR}}$) transmission coefficients are frequency-independent and equal to 1/2. While the phases of co-polar transmission coefficients are zero at any frequency, the cross-polar transmission coefficients have equal but frequency-dependent phases. Though a certain frequency behavior of $\phi_y$ is determined by a unit cell configuration, the above discussed phases of the co- and cross-polar transmission coefficients result in CP-to-LP conversion in transmission. Importantly, this function is only achievable for the considered self-complementary MSs with anisotropic patterns and translational symmetry.  Indeed, by substituting (\ref{TransmissionCPFinal}) into (\ref{TransmissionCP1}) we obtain
\begin{equation}\label{OutputLinear}
 \begin{gathered}
 \begin{pmatrix}
      E_{x}^{\text{tra}} \\
      E_{y}^{\text{tra}}
 \end{pmatrix} =
\overline{\overline{M}}^{-1}\cdot \overline{\overline{T}}_{\text{CP}} \cdot
\begin{pmatrix}
      E_{\text{R}}^{\text{inc}}\\
      E_{\text{L}}^{\text{inc}} \\
 \end{pmatrix} = \\
 =
\begin{pmatrix}
      {\frac{1}{\sqrt{2}}\exp{\frac{\text{i}\phi}{2}}\cos\frac{\phi}{2}\left(E_{\text{R}}^{\text{inc}}+E_{\text{L}}^{\text{inc}}\right)} \\
      {\frac{1}{\sqrt{2}}\exp{\frac{\text{i}\phi}{2}}\sin\frac{\phi}{2}\left(E_{\text{R}}^{\text{inc}}-E_{\text{L}}^{\text{inc}}\right)}
 \end{pmatrix}, 
 \end{gathered}
\end{equation}
where $\phi$=$\pi+2\phi_{y}$ is the phase difference between the cross-polar and co-polar transmission coefficients.
From (\ref{OutputLinear}) one may ensure that when the MS is illuminated by a circularly polarized wave, the transmitted one has purely linear polarization with the magnitude $|\textbf{E}^{\text{tra}}|=\frac{1}{\sqrt{2}} |\textbf{E}^{\text{inc}}|$. The angle $\alpha$ between the polarization vector of the transmitted wave and $x$-axis equals $\phi/2$ when the incident wave is RHCP and $-\phi/2$ when it is LHCP. In both cases, the phase of the transmitted wave is $\phi/2$. It is easy to demonstrate that the reflected wave is also linearly polarized with $|\textbf{E}^{\text{ref}}|=\frac{1}{\sqrt{2}} |\textbf{E}^{\text{inc}}|$. The phase shift between the transmitted and reflected  waves is always $90^{\circ}$, while their polarization vectors are orthogonal.  

\section{Simulation and \\Experiment}

To verify the theoretically predicted electromagnetic properties of anisotropic self-complementary metasurfaces illuminated with circularly polarized plane waves, in this section we introduce the MS consisting of complementary patches and apertures and study it numerically and experimentally in the frequency range of $230$$-$$540$ GHz. The geometry of the structure and the problem is shown in Fig.~\ref{Geometry}.    
\begin{figure*}[h!] 
  \centering
  \includegraphics[width=1\textwidth]{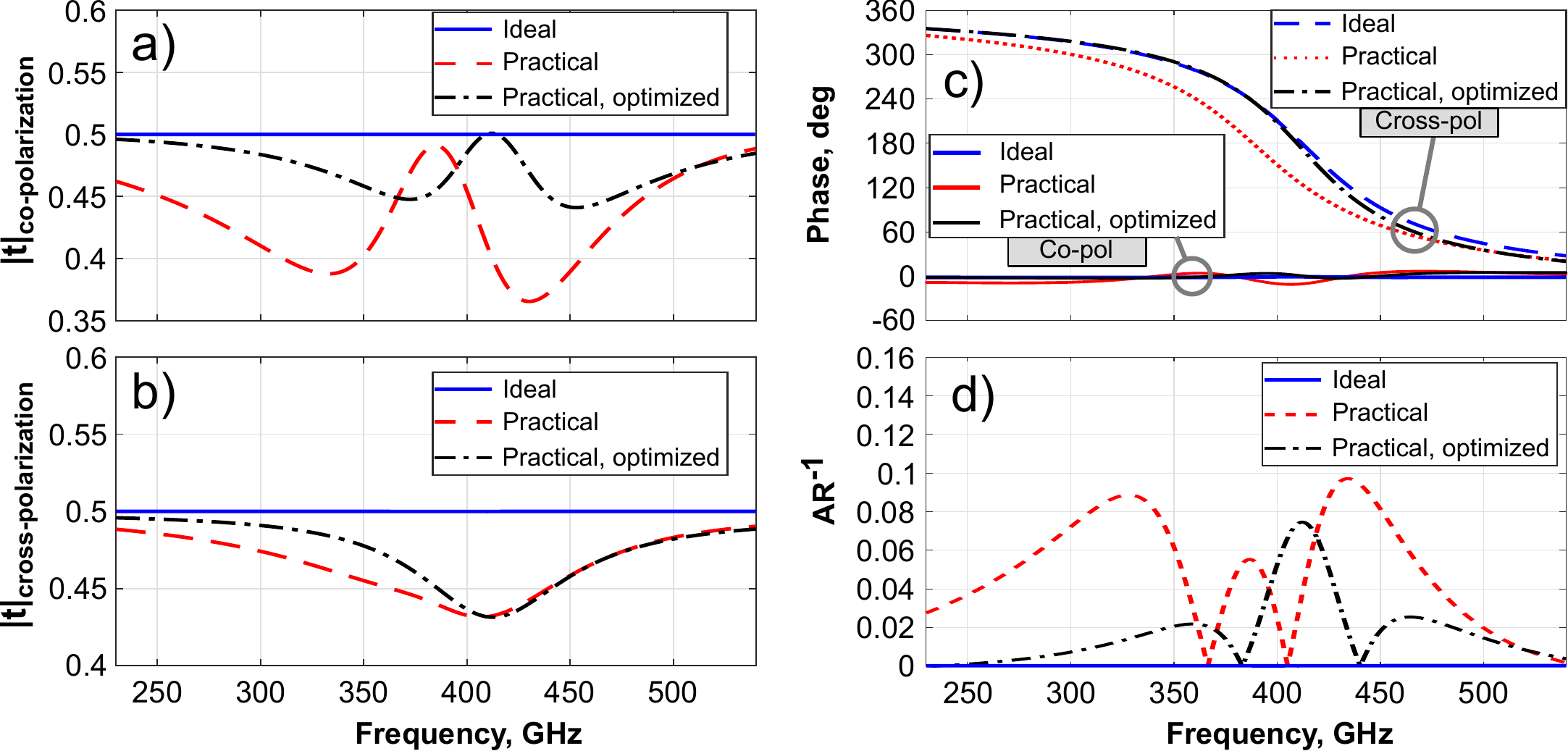} 
  \caption{Simulated transmission coefficients for the basis of circular polarizations: (a) magnitude of co-polarization; (b) magnitude of cross-polarization; (c) phases of co- and cross-polarizations. (d) Inverse axial ratio of the transmitted wave (RHCP illumination).}
  \label{t_sim}
\end{figure*}
From the formulas derived in the Section \ref{Th}, it follows that any anisotropic and loss-less infinitely thin self-complementary FSS converts the incident circular polarization into the linear one. The transmitted linearly-polarized wave has a frequency-constant magnitude and a polarization angle, which gradually depends on frequency. To confirm these properties, we numerically calculated the transmission and reflection coefficients of the ideal MS of patches and apertures patterned in a perfectly-conducting and infinitely thin metal sheet. The MS was illuminated with a right-handed circular polarization. The MS unit cell parameters were chosen from the condition of placing the MS resonance at 410 GHz: lateral periodicities in $x$- and $y$-direction $P_x = 170$ {\textmu}m and $P_y = 280$ {\textmu}m; widths of strips and gaps $W_{x1}=W_{x2}=15$ {\textmu}m and $W_{y1}=W_{y2}=7.5$ {\textmu}m.
The calculated transmission coefficients of the ideal MS for co- (RHCP) and cross- (LHCP) polarizations are plotted in Fig.~\ref{t_sim}(a-c) with blue solid curves.

Indeed, from Fig.~\ref{t_sim}(a-b) it is seen that the magnitudes of both co- and cross-polar transmission coefficients are equal to 1/2 in the whole frequency range. Fig.~\ref{t_sim}(c) shows the phases of the transmission coefficients. The co-polarization phase equals zero at all frequencies, while the phase of the cross-polar transmission coefficient is frequency-dependent and taking on the values from $20^\circ$ to $335^\circ$.
To ensure that the transmitted wave is linearly polarized, we plotted the inverse axial ratio (AR$^{-1}$) in Fig.~\ref{t_sim}(d) with the blue solid curve. Its value equals zero that means perfect CP-to-LP conversion.  

\begin{figure}[h]
  \centering
  \includegraphics[width=0.9\linewidth]{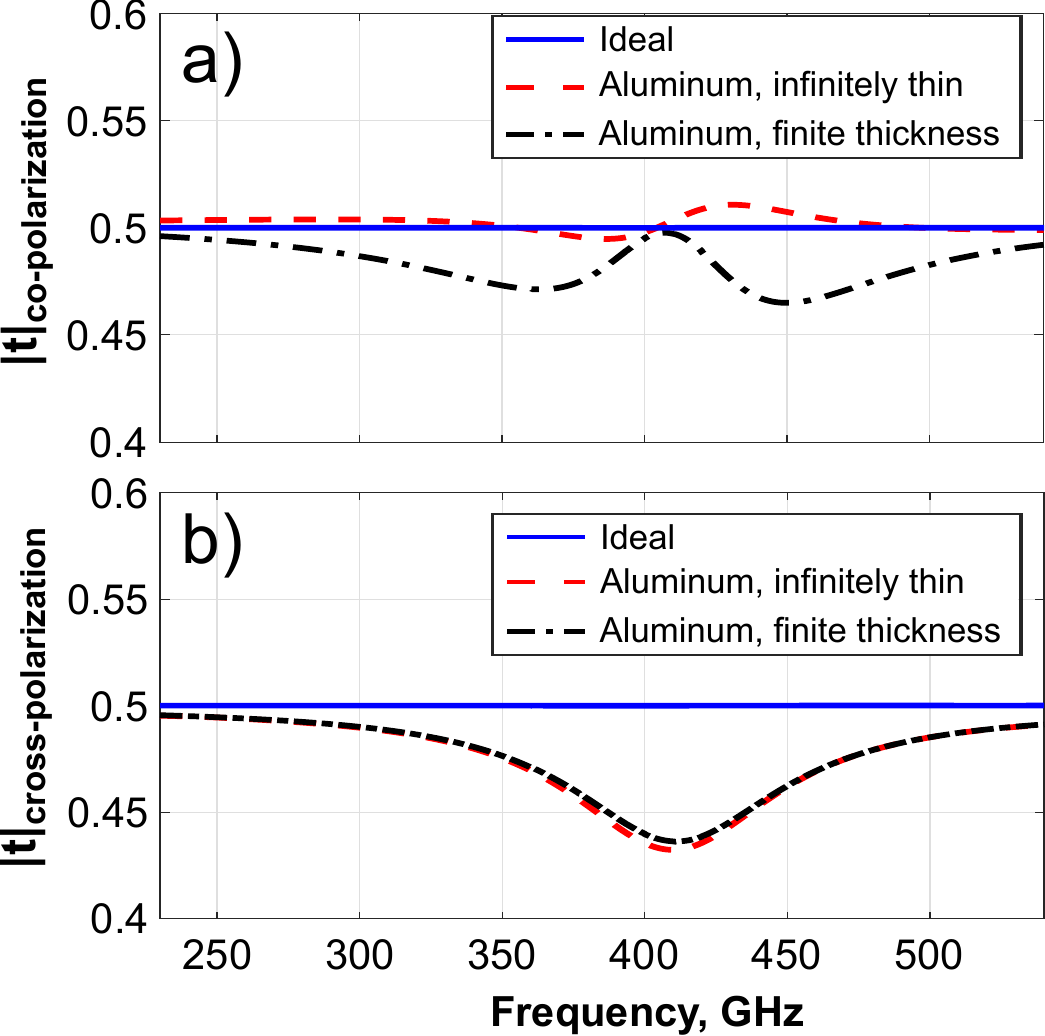} 
  \caption{Effect of losses and finite thickness of aluminum pattern on calculated transmission coefficients for circular polarizations: (a) co-polarization; (b) cross-polarization.}
  \label{Losses}
\end{figure}

At the next step, we have demonstrated the feasibility of the considered MS in the sub-terahertz range by studying the effects of lossy aluminum as a material of the metalization pattern and a realistic dielectric substrate.
First, we simulated a free-standing and infinitely thin MS made of lossy aluminum. The latter was modeled as a medium with the electric conductivity of 1.5E7 S/m whose dispersion in the frequency range of interests was neglected\cite{KuznetsovSciRep}. The simulation results are depicted in Fig.~\ref{Losses} with the red dashed curves showing that both co- and cross-polar transmission coefficients slightly differ from those of the ideal MS. 
We found that the magnitude of the co-polar transmission coefficient remains very close to 0.5 (the difference from the ideal case does not exceed 2\%) while the magnitude of the cross-polar transmission coefficient exhibits frequency variation within 12\%. Therefore, the metal losses substantially affects only the cross-polarization component of the transmitted wave.
In addition, the effect of a finite thickness of the aluminum pattern was studied. We considered a free-standing sheet made of lossy aluminum with the thickness of $0.4$ {\textmu}m  instead of zero thickness in the previous case. The numerical results are plotted in Fig.~\ref{Losses} with black dot-dash lines. It can be seen that increasing the thickness of the metal pattern only affects the co-polar transmission.

\begin{figure}[t]
  \centering
  \includegraphics[width=0.4\textwidth]{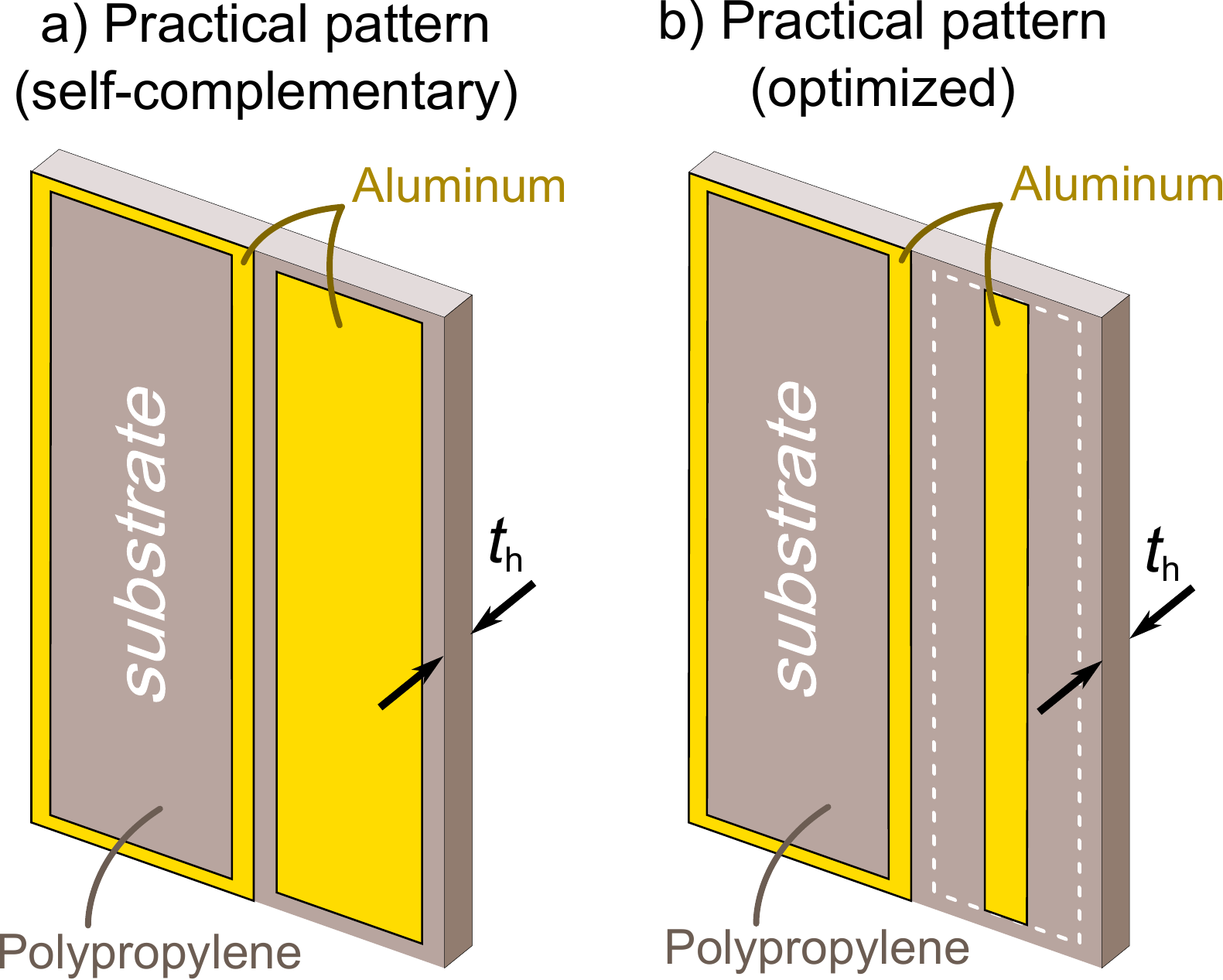} 
  \caption{Unit cells of practical metasurfaces of patches and apertures with self-complementary (a) and quasi-self-complementary (optimized) (b) patterns.}
  \label{Unit-cell}
\end{figure}

To consider further a practical realization of our self-complementary MS, it is to be noted that this kind of MSs inherently necessitates employing a dielectric substrate to mechanically support the MS pattern. For this reason both the effects of a finite thickness lossy metalization and a dielectric substrate were jointly studied. The $0.4$-{\textmu}m-thick aluminum pattern was backed by a $10$ {\textmu}m-thick polypropylene film, exactly as in the experiment described below. The dielectric permittivity of polypropylene was taken from the work\cite{KuznetsovSciRep}: $\epsilon=2.25\cdot(1-\text{i}\cdot0.001)$. The simulation show that the presence of the substrate leads to a $66$ GHz red shift of the MS resonance relative to the substrate-free case ($410$ GHz). To compensate this shift, we reduced the period $P_y$ from $280$ to $245$ {\textmu}m. The unit cell of the simulated self-complementary MS is shown in Fig.~\ref{Unit-cell}(a). In comparison with the ideal MS (blue solid curves in Fig.~\ref{t_sim}(a), the practical substrate-backed MS has considerably different properties as indicated with red dash lines in the same figure.
Thus, the difference in the co-polar transmission coefficient magnitude from the ideal MS reaches 28\%. In contrast, the magnitude of the cross-polar transmission coefficient is almost unchanged when the substrate is present (compare the red dash line in Fig.~\ref{t_sim}(b) with the black dash-dot line in Fig.~\ref{Losses}(b)). The calculated transmission coefficients phases are shown in Fig.~\ref{t_sim}(c). The cross-polar transmission phase remains frequency-dependent and covers the range of $20^\circ$$-$$326^\circ$. At the same time, the co-polar transmission phase represented by the red dashed curve, takes some deviations from zero, while not exceeding $10^{\circ}$. The inverse axial ratio AR$^{-1}$ of the practical MS does not exceed 0.1 as can be seen in Fig.~\ref{t_sim}(d). 

\begin{figure*}[ht]
  \centering
  \includegraphics[width=0.9\textwidth]{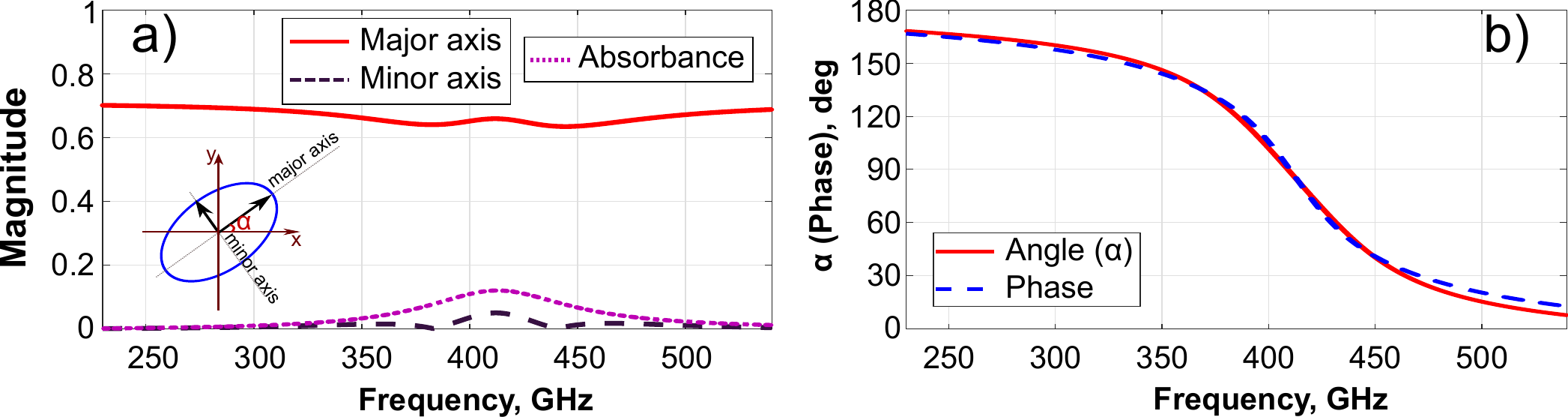} 
  \caption{Simulated characteristics of the wave passed through the optimized metasurface under right-handed circular illumination with a unit magnitude: (a) Absorbance (dot magenta) and lengths of the major (red solid) and minor (black dashed) axes of the polarization ellipse (shown in inset); (b) phase (blue dashed) and angle of the major axis (red solid).}
  \label{Output}
\end{figure*}

Previously, the presence of a substrate was found to violate self-complementarity of the structure\cite{Baena_self,Baena_zigzag}. This explains the frequency dependence of the transmission coefficients appeared when adding the substrate. However, the increased unit cell capacitance between the conductor edges caused by the substrate can be compensated via increasing the widths of the inter-edge gaps. In our case this optimization and subsequent recovery of the desired self-complementary properties was achieved by adjusting $W_{x2}$.
The optimized (quasi-self-complementary) MS unit cell with the gap width $W_{x2}=30$ {\textmu}m is shown in Fig.~\ref{Unit-cell}(b). The simulated results are plotted in Fig.~\ref{t_sim} with black dash-dot curves. By comparing these results with those of the substrate-free MS (black dash-dot curves in Fig.~\ref{Losses}), one can conclude that the performed optimization compensates the substrate effect. The optimized pattern provides the co-polar transmission coefficient magnitude higher than $0.46$ and the cross-polar magnitude higher than $0.44$ in the whole frequency range of interest. The cross-polar transmission phase as shown in Fig.~\ref{t_sim}(c) covers the range of $20^{\circ}$$-$$335^{\circ}$, while the co-polar phase deviates from zero within $4^{\circ}$. As a result, when the optimized MS is illuminated with a circularly polarized wave, the transmitted one has almost linear polarization with the inverse axial ratio AR$^{-1}$$\le 0.08$ (a relative level of the orthogonal linear polarization is below -21 dB in the whole frequency range). The numerically calculated properties of the transmitted wave are additionally illustrated in Fig.~\ref{Output}. In Fig.~\ref{Output}(a) the major and minor axes of the polarization ellipse for a unit magnitude of the incident RHCP wave are shown. In Fig.~\ref{Output}(b) the angle of the major axis $\alpha$ with respect to $x$-direction and the initial phase of the transmitted wave are plotted.
As expected, the length of the major axis is close to $1/\sqrt{2}$ (larger than $0.64$) and very frequency-stable, while the length of the minor axis is negligible (less than $0.04$). Interestingly, the angle $\alpha$ and phase of the transmitted wave are almost equal to each other and both depend on frequency covering the range from $8^{\circ}$ to $168^{\circ}$ as shown in Fig.~\ref{Output}(b). It is worth noting that the absorbance, defined as 1$-$0.5$\cdot$($|r_x|^2$+$|r_y|^2$+$|t_x|^2$+$|t_y|^2$), in our simulation does not exceed 13\% of the incident power. 

\begin{figure}[t]
  \centering
  \includegraphics[width=0.6\linewidth]{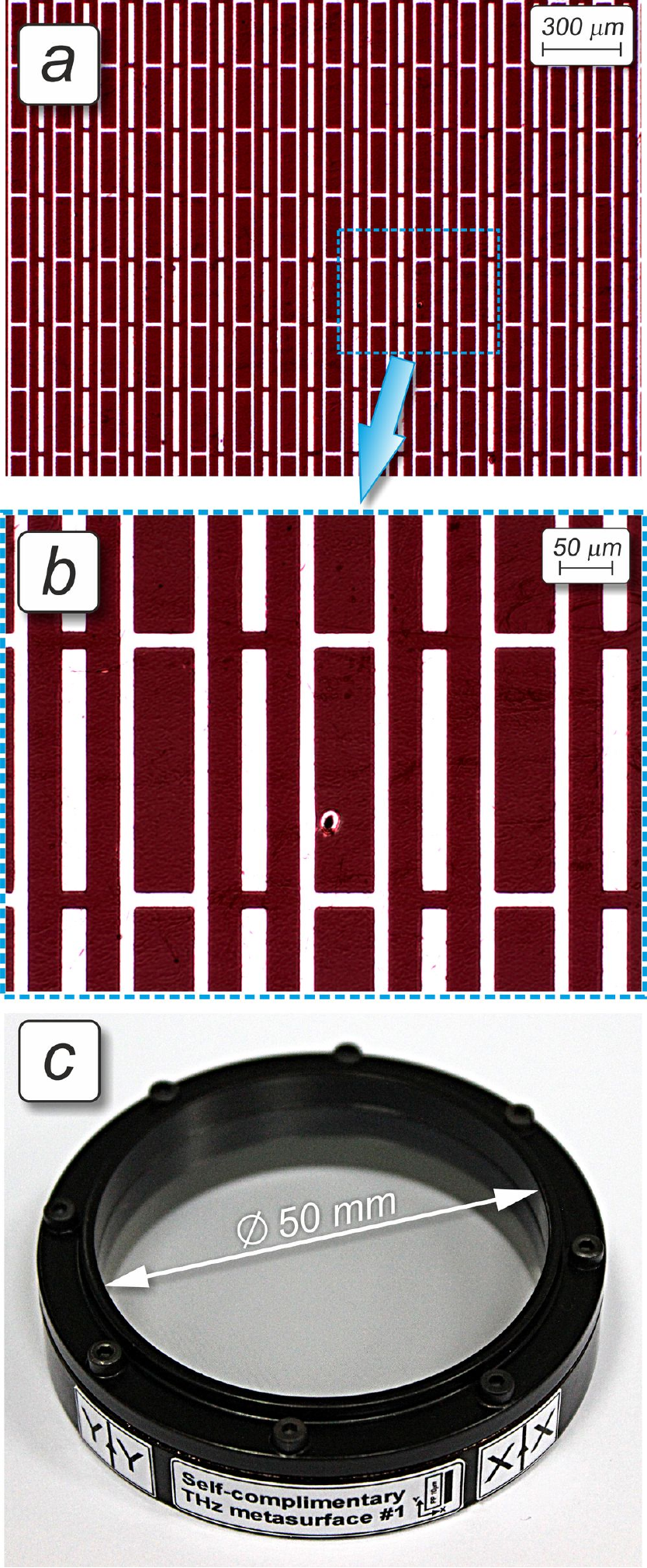} 
  \caption{Photographs of the lithographically fabricated MS pattern taken at different optical magnifications (a,b) and appearance of the completed structure mounted on the holder (c). Subfigure (b) illustrates the \textit{microhillock} defect (a bright spot in the center).}
  \label{Experimental_Sample}
\end{figure}

In order to verify further the theoretical and numerical results, the optimized MS was lithographically fabricated and its complex transmission coefficients were measured in the same frequency range as in the simulations (see Methods Section for details). The photographs of the fabricated MS sample are shown in Fig~\ref{Experimental_Sample}.

\begin{figure*}[th]
  \centering
  \includegraphics[width=0.9\textwidth]{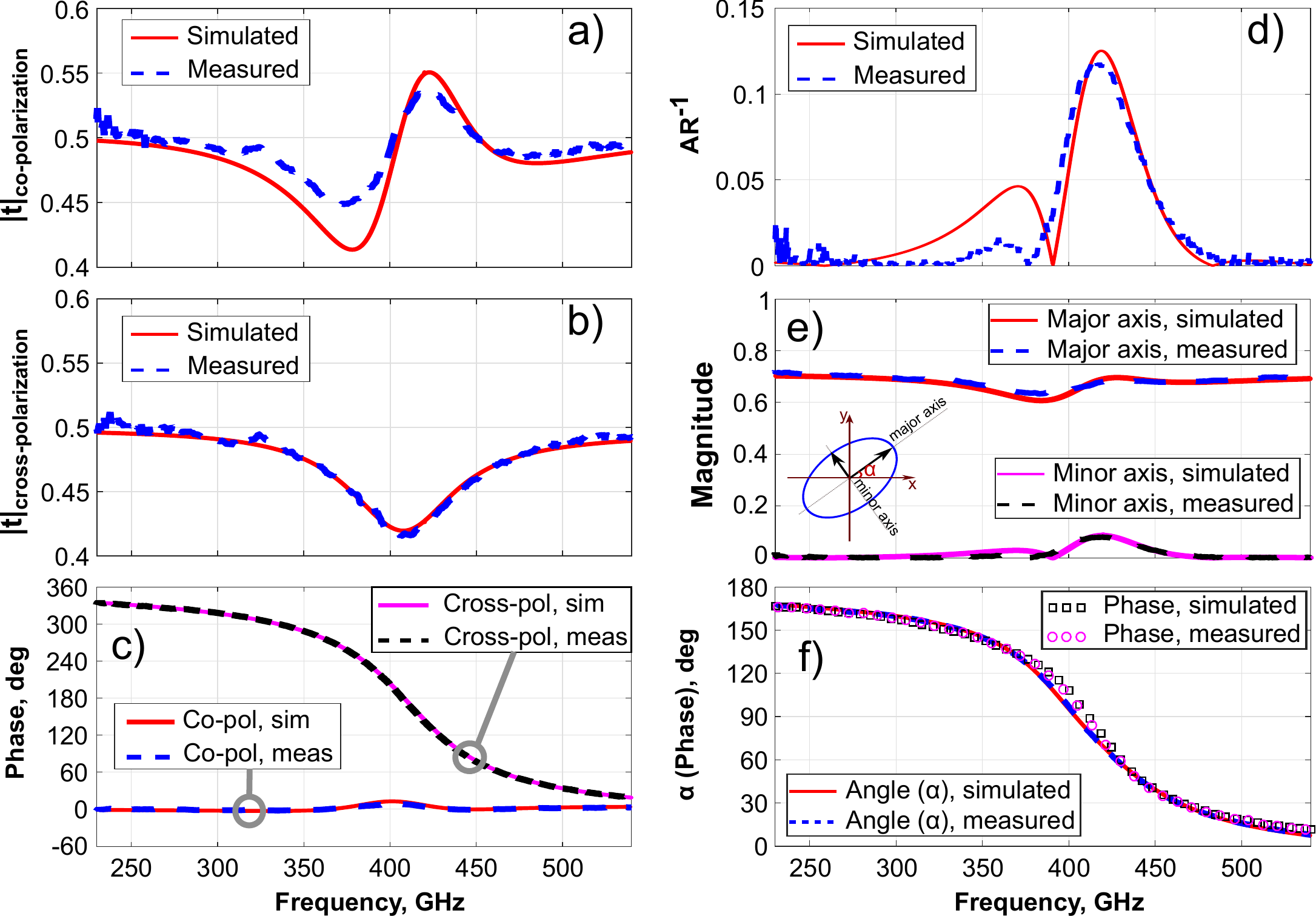} 
  \caption{Comparison of simulated and measured characteristics of the optimized MS under circularly-polarized illumination: (a-c) magnitudes and phases of transmission coefficients referred to the circular polarizations basis; (d-f) properties of the transmitted linearly polarized wave. Inset in subfigure (e) shows orientation of the polarization ellipse.}
  \label{t_meas}
\end{figure*}

The measured co- and cross-polar transmission coefficients in the basis of circular polarizations (blue dashed curves) are collated in Fig.~\ref{t_meas}(a-c). For comparison, additional numerical simulations were performed for the optimized MS taking into account some deviations in MS dimensions from the nominal values appeared in fabrication (see Methods Section). The simulation results are shown in Fig.~\ref{t_meas}(a-c) with red solid curves. The presented data demonstrate very good agreement between measurements and simulations, which predict the linear polarization state for the transmitted wave. To prove this, the polarization ellipse parameters of the transmitted beam derived from the experimental data are shown and compared to simulations in Fig.~\ref{t_meas}(d-f). From the measurements results one can see that the transmitted wave is mostly linearly polarized with a cross-polarization level below $0.08$. The transmission coefficient from RHCP to the main linear polarization is better than $0.61$, while being very stable within the whole range of frequencies. The polarization vector of the transmitted wave gradually changes with frequency from $8^{\circ}$ to $167^{\circ}$ with respect to the $x-$direction.
Therefore, our experiment confirmed with high accuracy all the theoretical and numerical predictions.
%
%The measured data is in a good agreement with the simulations. It worth to note that considered metasurface demonstrates a high stability of the characteristics with respect to the manufacturing errors.    
%

\section{\uppercase{Conclusion}}
In this work, we have clearly demonstrated that anisotropic self-complementary metasurfaces with translational symmetry of unit cells illuminated by a normally incident circularly polarized plane wave transmits the linear polarization with an almost frequency-constant magnitude and the polarization vector, gradually rotating with frequency. As was concluded from analytical and numerical results, the transmitted linear polarization originates from interference of two transmitted circularly polarized waves: with co- and cross-polarizations relative to the incident one. Indeed, the latter were theoretically and experimentally shown to be of almost constant magnitudes, but having a frequency-dependent phase difference resulted from the symmetry of self-complementary patterns. The THz characterization of the fabricated MS demonstrated good concordance with theoretical predictions. 

It can be concluded that the proposed terahertz MS consisting of patches and apertures may operate as an optically-thin CP-to-LP polarization converter with the opportunity to control the polarization angle of the transmitted wave by changing frequency. On the other hand, the MS produces in transmission two equally strong circularly polarized waves with opposite handedness. Interestingly, the complex transmission coefficient of self-complementary MSs for the circular co-polarization does not depend on frequency. And in fact, it is also independent on the unit cell geometry that repeats conclusion of the work\cite{Nakata}. However, the phase of the cross-polarization depends on frequency and on the unit cell geometry. Consequently, the cross-polarized transmitted beam may be controlled regardless of the co-polarized one. For instance, a non-uniform and anisotropic self-complementary MS can be applied in broadband and thin power splitters with output circularly polarized beams having opposite handedness but frequency-stable intensities.

\section{\uppercase{Methods}}

\subsection{Numerical simulation}
The full-wave numerical simulations of the MSs were carried out in commercial electromagnetic software CST Microwave Studio 2017 with a Frequency Domain Solver. The Bloch-Floquet ports were set from both sides of a unit cell in $z$-direction together with the Unit Cell boundary conditions applied in $x$- and $y$-directions. The complex transmission coefficients in the circular polarization basis were calculated in the frequency range from 230 to 540 GHz. Since the periodicity in both directions was smaller than a half-wavelength referred to the highest frequency, only two zero-order Bloch-Floquet modes were taken into consideration.

\subsection{Fabrication and measurement}
To fabricate the experimental prototype of the optimized MS the technology of contact photolithography (CPhL) was employed \cite{Kuznetsov2010,Navarro2011}. We specifically adapted CPhL to flexible solid thin-film polypropylene (PP) substrates, whose industrial production does not allow obtaining a liquid phase suitable for posterior film deposition via spin coating. At the initial fabrication stage a bare $10$ {\textmu}m thick PP film from GoodFellow manufacturer was metalized from one side by sputtering a $0.4$ {\textmu}m thick aluminum (Al) layer over the area of $60\times 60$ mm using a vacuum thermal deposition method. Prior to sputtering, the PP film was treated with a glow discharge in oxygen atmosphere to improve adhesion of Al to PP. The MS pattern was fabricated afterwards by dint of CPhL, as described in \cite{Kuznetsov2010,Navarro2011}. At the final fabrication stage the PP-backed MS was tightened onto an annular Al holder with the optical aperture diameter of 50 mm matched to the experimental set-up used for spectral characterization.
The micrographs of the fabricated MS pattern and the external appearance of the MS prototype are shown in Fig.~\ref{Experimental_Sample}. It is worth noting that, despite inherent surface roughness (microgranularity) of commercial PP films and presence of isolated \textit{microhillock} defects (see Fig.~\ref{Experimental_Sample}(b)) which cause local metallization flaking and/or pattern deformation \cite{Navarro2011}, the average quality of the fabricated MS pattern is estimated to be high: the typical dimension deviation from nominal values lies within $\pm 0.2$ um. Such a technological tolerance is acceptable for implementing high-performance PP-based structures operating in the THz band. It should be noted that due to stretching the PP film after its tightening onto the Al holder, the metasurface dimensions slightly changed. These changes lie within 1\% relative to the nominal (designed) values and the dimensions were measured to be: $P_x = 171.5$ {\textmu}m, $P_y = 247.3$ {\textmu}m, $W_{x1}= 14.8$ {\textmu}m, $W_{y1}= 7.4$ {\textmu}m, $W_{x2}= 30.9$ {\textmu}m and $W_{y2}= 8.2$ {\textmu}m.     

\begin{figure*}
  \centering
  \includegraphics[width=0.65\linewidth]{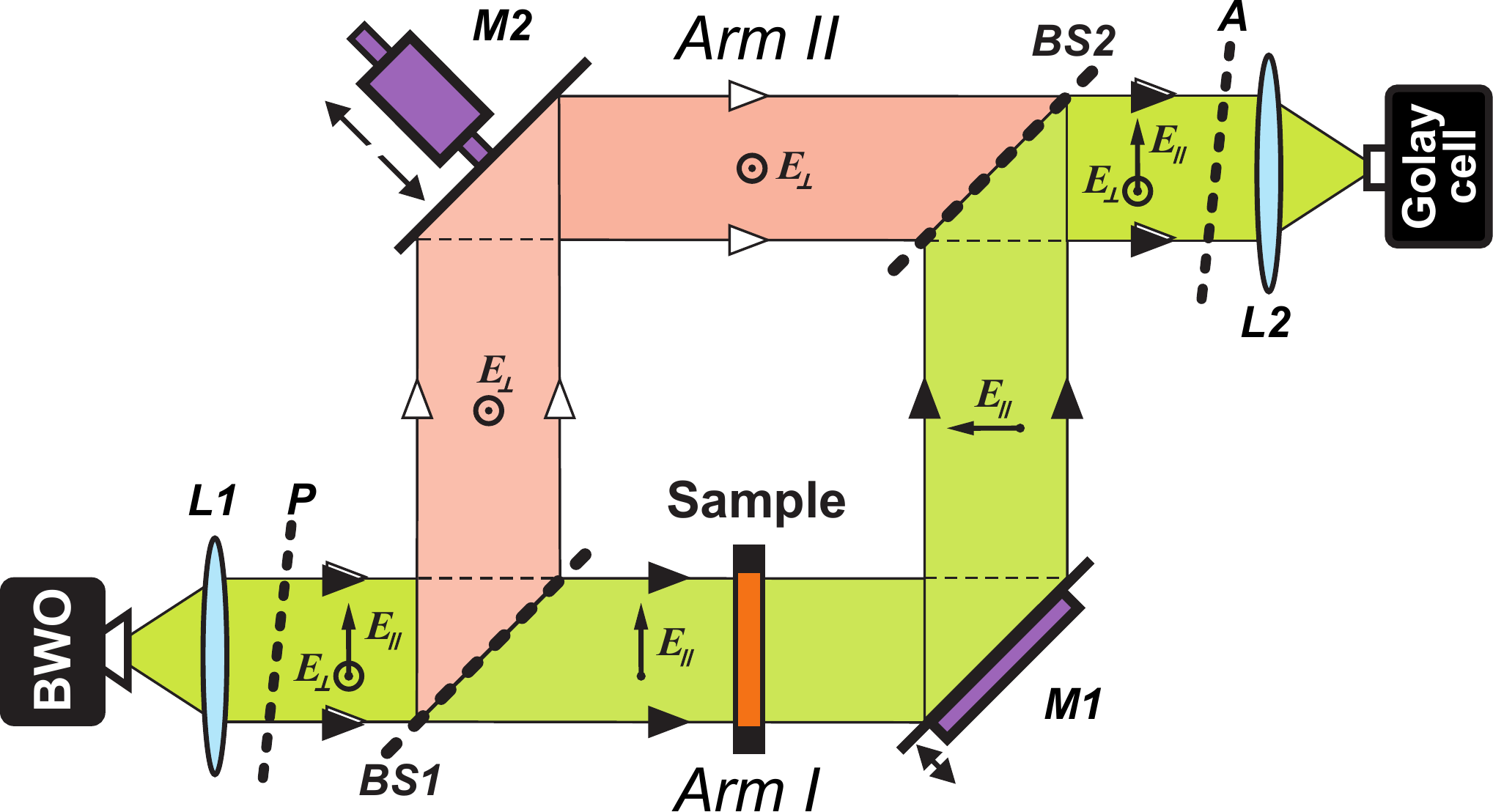} 
  \caption{Layout of the BWO-spectrometer in polarizing Mach-Zehnder interferometer arrangement used for measuring complex transmission of MS samples. $L1$ $\&$ $L2$ $-$ lenses, $P$ $-$ wire grid polarizer, $A$ $-$ wire grid analyzer, $BS1$ $\&$ $BS2$ $-$ wire grid beam splitters, $M1$ $-$ vibrating mirror, $M2$ $-$ movable mirror. The beams in measurement $Arm$ $\textrm{I}$ and reference $Arm$ $\textrm{II}$ have mutually orthogonal polarizations (designated by  vectors $E_{||}$ and $E_{{\bot }}$ respectively).}
  \label{Experimental_scheme}
\end{figure*}

To measure the complex transmission coefficient $S_{21}$ of the fabricated sample in the range of 230-540 GHz, a subterahertz quasi-optical backward wave oscillator (BWO) spectrometer developed by the Prokhorov General Physics Institute (Moscow, Russia) \cite{book:Kozlov} was used. This frequency-domain (continuous-wave) instrument exploits the arrangement of the dual-path polarizing Mach-Zehnder interferometer shown in Fig.~\ref{Experimental_scheme}. In the presented arrangement the coherent monochromatic THz radiation from a wavelength-tunable BWO passed through the wavefront-correcting polyethylene lens $L1$ and the wire grid polarizer $P$ is emitted into free space as a diffraction-limited Gaussian beam. The beam is further split with a wire grid beam splitter $BS1$ onto two orthogonally linear polarized THz beams propagating in the different arms of the interferometer, wherein the examined sample is placed in the measurement $Arm$ $\textrm{I}$. Owing to a beam splitter $BS2$, the beams gather at the analyzer $A$ and then interfere in the focal plane of the lens $L2$. The interference signal is quantified with a Golay cell used as a broadband THz detector combined with a lock-in amplifier operating at the modulation frequency of 23 Hz. Note, the optimal orientation of wires in the wire grid elements implies that the polarizer $P$ and the analyzer $A$ are mutually crossed and oriented at $45^{\circ}$ relative the beam splitters $BS1$ and $BS2$, whose wires are also mutually orthogonal (vertical and horizontal, respectively).
In amplitude measurements, only the measurement $Arm$ $\textrm{I}$ is utilized. In this case a mechanical obturator (chopper) is applied for 23 Hz shuttering, as required in the lock-in detection scheme, and the value of $|S_{21}|^2$ is determined as a ratio of the intensity $I_{\text{sample}}$ for the THz beam passed through the sample to the intensity $I_{\text{free}}$ registered without it: 
\begin{equation}
\label{transmittance}
|S_{21}|^2=\frac{I_{\text{sample}}}{I_{\text{free}}},
\end{equation}
When measuring the transmission phase ($\phi_{\text{meas}}$), both interferometric arms are involved. During spectral measurements, the optical lengths of the $Arm$ $\textrm{I}$ and the $Arm$ $\textrm{II}$ are adjusted to maintain a zero-order interference minimum that is accomplished via changing a position of the movable mirror $M2$, which is combined with a computer-controlled longitudinal translation stage and has a positioning accuracy of 0.5 $\mu$m. In this case, the lock-in detection is realized not with a chopper but through vibrating the surface of the mirror $M1$, which is attached to a loudspeaker membrane oscillating at 23 Hz. In the measurement procedure, when varying the wavelength of the THz beam emitted by the BWO, the value of $\phi_{meas}$ is retrieved by spectrometer software from a difference in the automatically recorded positions $X_{\text{sample}}$ and $X_{\text{free}}$ of the mirror $M2$ correspondingly obtained when the sample is placed in the measurement $Arm$ $\textrm{I}$ and when it is removed from the optical path:
\begin{equation}
\label{measured_phase}
\phi_{\text{meas}}=-\frac{2\pi}{\lambda}(X_{\text{sample}}-X_{\text{free}}+d),
\end{equation}
where $d$ is the sample thickness. Note, the regime of zero-order interference is recommended as optimal since it eliminates frequency dispersion in the mirror position in absence of the sample. 

Once the complex transmission coefficients in the linear polarizations basis are measured, the transmission coefficients in the basis of circular polarizations can be evaluated using (\ref{TransmissionCP2}).
The axial ratio is calculated as a ratio between the major and minor axes of the polarization ellipse of the transmitted wave and is expressed through the complex magnitudes of transmitted electric field components $E_x$, $E_y$ as follows \cite{Baena_zigzag}:
\begin{equation}
\label{AR}
\text{AR}=\sqrt{\frac{|E_{x}|^2+|E_{y}|^2+|E_{x}^2+E_{y}^2|}{|E_{x}|^2+|E_{y}|^2-|E_{x}^2+E_{y}^2|}},
\end{equation}

\bibliography{Bibl}

\end{document}